\documentclass{article}

\usepackage{authblk}

\usepackage[latin1]{inputenc}
\usepackage{amsmath}
\usepackage{amssymb}
\usepackage{amsthm}
\usepackage{amsfonts}
\usepackage{abstract}
\usepackage{mathrsfs}
\usepackage{hyperref}
\usepackage{mathtools}
\usepackage{bm}
\usepackage{bbm}
\usepackage[english]{babel}
\usepackage{color}
\usepackage{graphicx}

\title{Comment on:  Maximal violation of Bell inequalities by position measurements}

\author[1]{Detlef D\"urr\thanks{duerr@mathematik.uni-muenchen.de}}
\author[2]{Sheldon Goldstein\thanks{oldstein@math.rutgers.edu}}
\author[3]{Tim Maudlin\thanks{twm3@nyu.edu}}
\author[1]{Robin Schlenga\thanks{robin.schlenga@physik.uni-muenchen.de}}
\author[4]{Daniel V. Tausk\thanks{tausk@ime.usp.br}}
\author[2]{Roderich Tumulka\thanks{tumulka@math.rutgers.edu}}
\author[5]{Nino Zanghì\thanks{Nino.Zanghi@ge.infn.it}}
\affil[1]{Mathematisches Institut der LMU, Theresienstr. 39, 80333
M\"unchen, Germany}
\affil[2]{Department of Mathematics, Hill Center,
Rutgers University, Piscataway,
USA}
\affil[3]{Department of Philosophy, NYU, New York, USA}
\affil[4]{Department of Mathematics, University of S\~ao Paulo, S\~ao Paulo, Brazil}
\affil[5]{Department of Physics, University of Genova, Genova, Italy}

\date{}

\begin{document}
\maketitle

\section*{}

In the 2010 \emph{research highlight}  article ``Maximal violation of Bell inequalities by position
measurements'' of the \textsc{Journal of Mathematical Physics},
Kiukas and Werner  \cite{KiukasWerner} critique  the empirical adequacy of Bohmian mechanics as an explanation of the phenomena of non-relativistic quantum mechanics. They observe  that (1)   two-time correlations of unmeasured positions in Bohmian mechanics  are different from those of the measured ones. Because of this they claim that (2) even  a position measurement at a single time need not reveal the Bohmian position at that time. They also write that (3) a reconciliation between Bohmian mechanics and the experimental predictions of orthodox quantum theory can be achieved only at the expense of adopting ``a strongly contextual view."

(1) has been observed many times before.  It has been discussed in great detail in \cite{BerndlPRA}, where it is also shown that the  statistics of measured positions for Bohmian mechanics agree with the quantum mechanical ones. (2) is simply false.
(3), while true, is so misleading as to be morally false. Let us explain.

We first address (3). What is meant by a contextual view is this: that different experimental procedures for measuring the same quantum observable must be regarded as physically inequivalent, at least potentially. As Bohr \cite{Bohr} has insisted, it is necessary in quantum mechanics to take the full experimental arrangement into account.

Measurements of quantum observables in Bohmian mechanics are indeed contextual: though Bohmian mechanics is deterministic, different experiments for measuring the very same quantum observable can lead to different measured values even when the experiments are performed on systems with the very same initial conditions. The issue of contextuality in Bohmian mechanics is discussed in detail in \cite{RoleOfOperators}, where the point is made that the contextuality in Bohmian mechanics, if not a complete triviality, is not nearly the dramatic innovation that it is sometimes suggested to be.

 For our purposes here, the crucial point about contextuality in Bohmian mechanics is this: Whether one likes it or not, contextuality in Bohmian mechanics is simply a fact, and not a point of view that one may or may not adopt. It is not an  add-on to Bohmian mechanics, adopted in order to restore compatibility with the predictions of orthodox quantum theory. Rather it is a feature of the original theory itself. One could, we suppose, consider a hidden-variables alternative to quantum mechanics that is non-contextual. That theory would not be Bohmian mechanics. And it would be incompatible with the predictions of quantum mechanics---as Bell's inequality shows.

We now turn to (2), and to how it can be compatible with (1), appearances to the contrary notwithstanding. Perhaps the simplest way to appreciate what is going on is in terms of the conditional wave function, an extremely useful notion in Bohmian mechanics that plays the role of wave function of a system of interest, usually a subsystem of a larger system---consisting of a system and  its environment or a system and apparatus.

For simplicity, let us consider the case of particles without spin. While the subsystem is usually entangled with its environment, so that in orthodox quantum theory it might not seem to have its own wave function but only a density matrix (the reduced density matrix), in Bohmian mechanics it does: its conditional wave function, given by

$$\psi( x)\sim\Psi( x, Y),$$
where $x$ represents the generic configuration of the system and $Y$ the actual configuration of its environment, with $\Psi$ the wave function of the combined system. It is easy to see that $\psi$ governs the evolution of the actual configuration $X$ of the subsystem in the usual Bohmian way, and that it collapses upon  measurement of the subsystem according to the usual quantum mechanical rules, with probabilities given by the Born rule.

In particular, when a position measurement is performed on one particle of a (sub)system, that system's  wave function $\psi$ and the  future evolution of its particles will normally be affected, so  that the future positions of its particles---and the results of future position measurements---will be different from what they would have been had the first  position measurement not taken place.   This is true regardless of whether the subsystem consists of a single particle, or involves two or more possibly entangled particles. In the latter case the first measurement on one of the particles could change the wave function of a distant particle, and hence its measured position in the future. Nonetheless, each of the position measurements would find the measured particle at its actual position. It's just that that actual position, found in a measurement subsequent to the first, might well be different from what it would have been had no first measurement been taken.

The uncertainty principle, valid in Bohmian mechanics as well as in orthodox quantum mechanics, provides a simple example involving just a single free  particle. Suppose an accurate position measurement is performed on that particle. This will transform its wave function to one with very narrow support and hence very large spread in momentum space. Thus a subsequent position measurement will be likely to yield much larger values than would be expected without the first measurement. This is true in Bohmian mechanics \cite{RoleOfOperators} exactly as it is in orthodox quantum theory.

Indeed, it would be entirely fair to say that Bohmian mechanics is the precise theory closest to what physicists have in mind when they use quantum mechanics---regardless of whether or not they realize this.

\end{document}